# Interpretable learning of voltage for electrode design of multivalent metal-ion batteries

Xiuying Zhang[1,2], Jun Zhou[3], Jing Lu[2,4,*], Lei Shen[5,*]

[1]Department of Physics, National University of Singapore, 2 Science Drive 3, Singapore 117542, Singapore.

[2]State Key Laboratory of Mesoscopic Physics and Department of Physics, Peking University, Beijing 100871, P. R. China.

[3]Institute of Materials Research & Engineering, A∗STAR (Agency for Science, Technology, and Research), 2 Fusionopolis Way, Innovis, Singapore 138634, Singapore.

[4]Collaborative Innovation Center of Quantum Matter, Beijing 100871, P. R. China

[5]Department of Mechanical Engineering, National University of Singapore, 9 Engineering Drive 1, Singapore 117542, Singapore.

[*]Corresponding author: jinglu@pku.edu.cn; shenlei@nus.edu.sg

**Abstract:** Deep learning (DL) has indeed emerged as a powerful tool for rapidly and accurately predicting materials properties from big data, such as the design of current commercial Li-ion batteries. However, its practical utility for multivalent metal-ion batteries (MIBs), the most promising future solution of large-scale energy storage, is limited due to the scarce MIB data availability and poor DL model interpretability. Here, we develop an interpretable DL model as an effective and accurate method for learning electrode voltages of multivalent MIBs (divalent magnesium, calcium, zinc, and trivalent aluminum) at small dataset limits (150~500). Using the experimental results as validation, our model is much more accurate than machine-learning models which usually are better than DL in the small dataset regime. Besides the high accuracy, our feature-engineering-free DL model is explainable, which automatically extracts the atom covalent radius as the most important feature for the voltage learning by visualizing vectors from the layers of the neural network. The presented model potentially accelerates the design and **optimization of multivalent MIB materials with fewer data and less domain-knowledge restriction, and is implemented into a publicly available online tool kit in** http://batteries.2dmatpedia.org/ **for the battery community.**

Lithium-ion batteries (LIBs) have shown unprecedented success as the major power source for transportation and as an energy storage solution for grid applications[1-3]. Nowadays, its relatively low energy density and the scarcity of Li raw materials are the main issues of LIBs for large-scale applications[4,5]. These issues call for high density, cheaper and sustainable alternatives to the present LIB technologies. Multivalent metal-ion batteries (MIBs), including $Mg^{2+}$, $Ca^{2+}$, $Zn^{2+}$, $Al^{3+}$, have the potential to meet this purpose, due to the relatively high abundance of these elements in the Earth's crust and high energy density[6].

Within the arena of big data, deep learning (DL) has emerged as a game-changing technique very recently, enabling numerous scientific applications in chemistry[7], mathematics[8], physics[9], and biology[10]. In materials science, several DL models have also been developed, such as AtomSets[11], SchNet[12], MatErials Graph Network (MEGNet)[13], Crystal Graph Convolutional Neural Network (CGCNN)[14-19], and have achieved great success in the applications, for instance, identifying degradation patterns of LIBs[20], designing solid-state LIBs[21], learning properties of ordered and disordered materials from multi-fidelity data[22], discovering stable lead-free hybrid organic-inorganic perovskites[23], screening 2D ferromagnets[24], discovering and designing electrocatalysts[25,26], mapping the crystal-structure phase[27], and designing material microstructures[28] It has been widely demonstrated that DL has much lower model errors than conventional machine learning (ML) models when dealing with big data. For example, the reported mean absolute error (MAE) of the deep neural network (DNN) model is lower than ML in predicting the volume change and voltage of LIB electrodes[29,30]. However, DL still cannot solve many problems in the field of batteries, especially the MIBs beyond lithium, due to insufficient data available. For example, Joshi et al. predicted the voltages of Na-ion battery electrodes with a DNN model, but the MAE is much higher than that of LIBs[29]. Moreover, due to being highly complex, DNN takes a long training time and is not explainable, which may not outperform shallow learning in solving some battery problems, especially for the design of batteries, in which explanations are in great demand. Therefore, the high-performance and interpretable DL model is of high need for predicting a variety of properties of multivalent MIBs from small data and then designing high-performing multivalent MIBs.

In this work, we take the voltage of multivalent MIBs as an example to demonstrate how an interpretable deep transfer-learning (TL) model (**Fig. 1**) can be used for exploration and design of electrode materials for battery applications, addressing the ML issues (low prediction accuracy and heavy feature-engineering dependence) and DL limitations (low interpretability and high big-data demand). It is worth noting that high-voltage electrode materials can enhance the voltage platform of batteries, which is the key component for high energy density MIBs and is generally used in the performance prediction of materials of battery electrodes[29-31]. We firstly train our DL models with relatively large data of the electrode voltage of LIBs (2000+ data) from Materials Project (MP)[10,32]. The MAE for LIBs is only 0.32 V. Nevertheless, using the same method, the MAE is significantly high for multivalent MIBs (up to 2.14 V) due to their small data (as low as 149). We, thus, integrate the TL technique [33,34], widely used to address less data restriction [35], into our DL models. It greatly reduces the MAEs for Zn-, Ca-, Mg-, and Al-ion batteries, for example, from 2.14 V down to only 0.47 V for the Zn-ion battery. To interpret our DL models, we perform the visualization of the similarity between the elements and local environments in different layers of the deep neural network. The DL models can automatically hierarchically extract key features and explain the different contributions of element groups in the periodic table to the corresponding electrode voltages. Our results show that the highly accurate and interpretable deep model could accelerate the discovery and design of electrode materials for multivalent MIBs and the development of the large-scale battery industry.

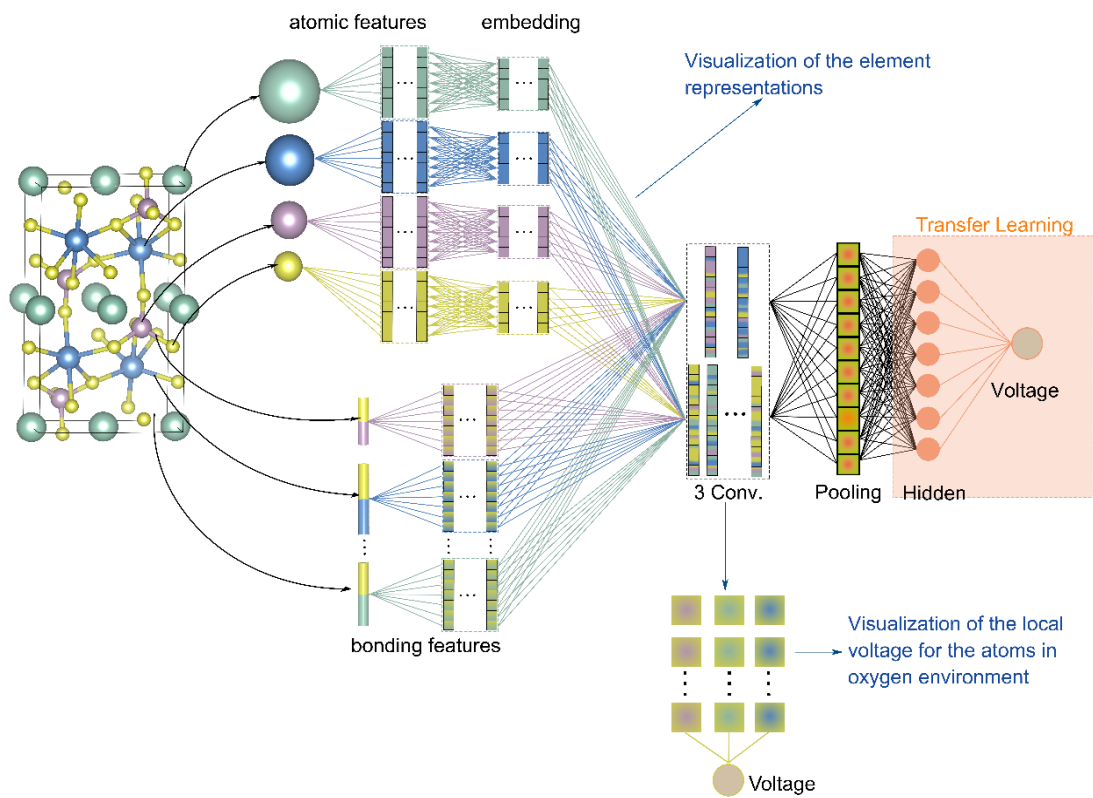

**Fig. 1 | Illustration of the interpretable deep neural network.** A crystal is converted and characterized by atomic and bonding vectors. The elemental and the local environment representations for visualization of the model are constructed from the embedding and the convolutional layers, respectively. The weights in the hidden layer and output layer are trained during transfer learning.

**Results and Discussion**

**Deep learning of Li-ion batteries.** 2190 samples of LIB electrode materials are used for model training and assessment after excluding the outliers and inconsistencies. The data (target data) firstly is randomly split into the training, validation, and test parts with the ratio of 0.8:0.1:0.1, respectively. The three parts show similar distributions (**Fig. 2a**, top), indicating a reasonable split. Then, we train and supervise our DL model (**Fig. 1**) with the training and validation data, respectively. The mean squared errors (MSEs) are chosen to be the loss function and MAEs are the evaluation metric. After fully training, the MAE of our DL model for predicting the voltages of LIBs is only 0.32 V, which is lower than the ML results (0.40 V). Our ML results are quite similar to the previous reports for LIBs[29,30]. The predicted voltages have similar distributions (**Fig. 2a**, right) with the target voltage (**Fig. 2a**, top) for all the training, validation, and test sets, and the points for the predicted voltages plotted against the target voltages are around the straight line of $y = x$ (**Fig. 2a**, dash line), which also indicates the high accuracy of our model. However, our DL model developed from LIBs (named Li-model) cannot be directly used for multivalent MIBs (quite large MAEs **Figs. 2b-2e**) because of the different properties between LIBs and multivalent MIBs. It's also inadvisable to train DL models from small data of multivalent MIBs. Furthermore, it is found that the MAEs of ML are similar to the Li-model (**Figs. 2b-2e**). Thus, neither DL nor ML is suitable for multivalent MIBs. This critical problem will be addressed in the late part of this paper.

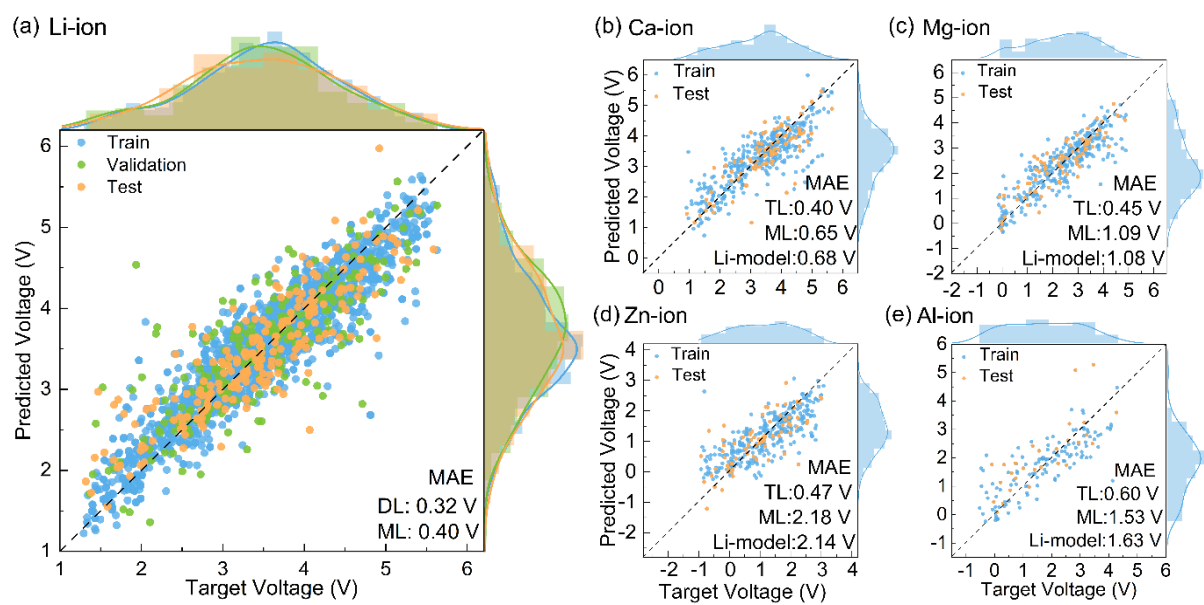

**Fig 2 | Plots of predicted voltage and target voltage for the metal-ion batteries.** The data points in **a** are predicted with the deep learning model and those in **b-e** are predicted with the transfer learning model. The model errors of the Li-, Ca-, Mg-, Zn-, and Al-ion battery. The histograms at the top and right show the distributions of predicted and target voltages, respectively. The black dashed line is the identity line ($y = x$) for reference. The TL, ML, and Li-model mean the transfer learning, machine learning, and model trained on LIB database only.

**Interpretability and visualization of the DL model.** Although the state-of-the-art DL models outperform ML models in the large dataset regime, they are generally viewed as black-box models due to the high complexity and are often achieved at the cost of interpretability. This is a major drawback of DL models for applications in which the interpretability of decisions is a critical prerequisite, such as new materials discovery and design[15,36,37]. Here, we visualize the embedding and convolutional layers of the deep neural network, separately[15] (see in **Fig. 1**) to interpret the contribution of underlying features to the electrode voltage prediction. Taking the LIB as a proof of concept, we first project the 64-dimensional elemental representation vectors on a two-dimensional (2D) plane constituted by the first two principal components, dimension 1 and 2, using the principal component analysis (PCA)[38,39]. **Figure 3a** shows the visualized element features automatically extracted from the embedded layers (**Fig. 1**). As can be seen, the elements can be clearly clustered into three groups according to the location of the corresponding elements in the periodic table. For instance, the first two group elements, alkali, and alkaline earth elements, locate at the upper right corner of the plot; the early transition metal (TM) elements locate at the left side; and the rest elements mainly distribute at the lower right side. Such a distribution indicates that the voltage of a crystal has a strong correlation with the group number of the constituent elements. Interestingly, we can observe a linear relationship between the covalent radius of the elements and the second principal component (dimension 2) (**Fig. 3a** inset). Thus, the covalent radius of the constituent elements of electrodes should be one of the important features for crystal voltage prediction. An ML-based random forest regression (RFR) model with 273 materials features[40,41] is also trained for analyzing the importance factor by the weight of features. The covalent radius is also found to be the most important feature with the importance ratio as high as 20% (**Fig. S1**). Such a result indicates the validation of the visualization analysis of our interpretable DL model.

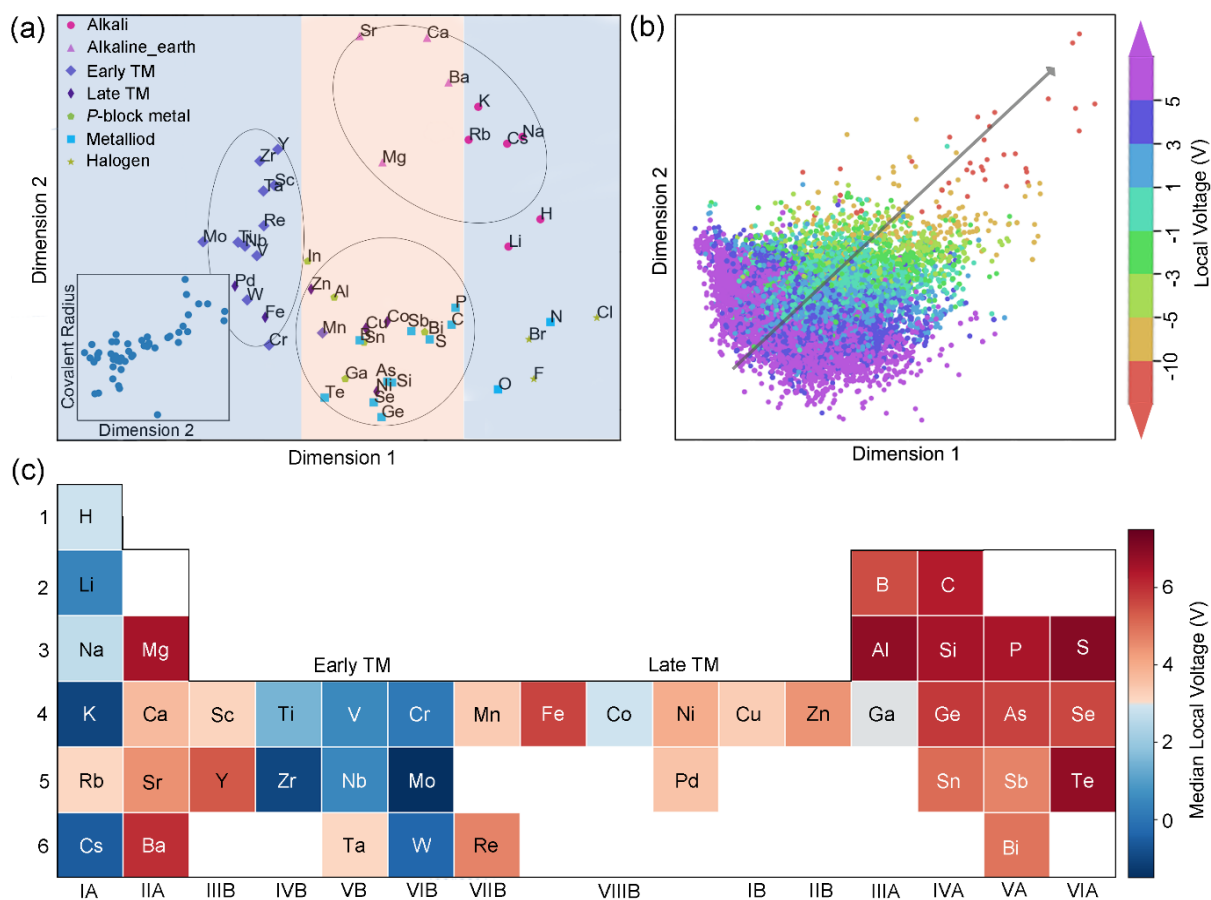

**Fig. 3 | Visualization of the DL model. a-b** The visualization of the two principal dimensions with PCA for the element representations and the local oxygen coordination environments representations, respectively. Dimension 1 and 2 in the plot constitute a plane in the corresponding vector space that can approximate the representations best. The points in **a** are coloured according to their elemental groups. **c**, the median value of the learned local voltages for each element in the oxygen coordination environments depicted in the elemental parodic table. The elements in **b** and **c** are labeled according to the type of the centre atoms in the oxygen coordination environments and the colours are coded with learned local voltages. The grey arrow in **b** indicates the change of local voltages.

Besides the visualization of the element representations in the neural network, four interpretable model can get a clear picture of the local environment from the electrode-voltage prediction. We next visualize the local environment (see more details in the Methods section), and take the oxygen coordination, at least two oxygen atoms around the central element, as the local environment representation because of the large number of oxides for battery electrodes. There are 23022 local oxygen coordination environments in total in our dataset. We first project all the oxygen local environments into a 2D plane using the PCA (**Fig. 3b**), and the elements are colored according to their local voltage which is calculated using Eq. (2) in **Methods**. It shows that the local voltage color changes along the diagonal direction (grey arrow) in the dimension 1- dimension 2 plot. Such a phenomenon indicates that both dimension 1 and dimension 2 have linear relationships with the local voltage, implying the local voltage information has been learned by our model. To surface such a relationship, we visualize the oxygen coordination environments with the t-distributed stochastic neighbor embedding (t-SNE) algorithm[42] (**Fig. S2**). It is found that the oxygen environments can be approximately grouped into three large groups according to the t-SNE distribution: i) the IA group with low voltages; ii) the main group with high voltages; and iii) the TM group with both low and high voltages. In order to validate such voltage performance extracted from the neural network, we calculate and project the median voltage ($V_i$) for each element (under the oxygen coordination environment) on the elemental periodic table as shown in **Fig. 3c**. As can be seen, it is quite similar to the t-SNE analysis, but with more details. The elements in the main group have high median local voltage except for the alkali metal elements which have low voltages. For the TMs, the early TMs mainly have the relatively low median local voltages, while the late TMs perform reversely. Interestingly, the PCA of the elemental features (**Fig. 3a**) also performs similar element-voltage relations with that of the local voltage of oxygen coordination environment analysis. For example, the elements with the low local voltages occupy the left and right parts (two regions shaded ion blue in **Fig. 3a**), while the middle area shaded in orange is mostly occupied by the elements with the high local voltages. This means that instead of extracting from the late convolutional layers (**Fig. 1**), the voltage information has already been learned in the early layer in our neural work. The oxygen coordination environment and the covalent radius shows a similar trend with respect to the element (**Fig. S3**), in accordance with the previous element representation analysis.

**Transfer learning of multivalent metal-ion batteries.** In this section, we will address the aforementioned problem that it is hard to get high-performance DL models for multivalent MIBs because of data scarcity, for example, only 149 Al-ion samples. Multivalent MIBs have the similar mechanism with LIBs, which allows us to reuse the pre-trained DL model on LIBs to multivalent MIBs. Here, we integrate the layer-freezing TL method[39], which would only fine tune the last two fully connected layers, into our deep neural network (see in **Fig. 1**) to predict the voltages of Mg-, Ca-, Zn-, and Al-ion battery electrodes. For comparison, the ML models and the DL model on LIBs (named Li-model) is also used on multivalent MIBs. **Figures 2b-2e** show the model error, MAE, for voltage prediction using the ML model, Li-model, and TL model. As can be seen, the Li-model and ML models have very similar MAEs which are much larger than those predicted by the TL model. It demonstrates that the prediction accuracy from small data has really been largely improved by the TL technique, especially for the Zn and Al-ion battery electrodes, whose voltage predictions have been significantly improved by 1.67 V and 1.03 V, respectively. Moreover, **Figs. 2b-2e** show that the predicted and the corresponding target voltage datasets have similar normal distributions, indicating that the datasets are reasonable, and the predictions are accurate. It is worth noting that the Li-model can provide a relatively good prediction for Na- and K-ion batteries (**Table S2**) because the Li-, Na-, and K-ions are all monovalent and having similar performances in electrodes[29]. However, our TL model still can further improve the performance in predicting monovalent MIBs (**Table S2**).

Finally, we validate our model by comparing it with the ML model and the available experimental results as shown in **Fig. 4**. As can be seen, our TL model outperforms the ML model in predicting the electrode voltage for multivalent Mg-, Ca-, Zn-, and Al-ion batteries. Therefore, the voltages predicted by the TL model are more reliable to guide experiments in finding appropriate electrode materials for improving the performance of future multivalent MIBs. To help the battery community, we thus implemented our model into a publicly available online tool kit that can be used for quickly pre-checking and screening the voltage of any novel electrode with the crystal structure only.

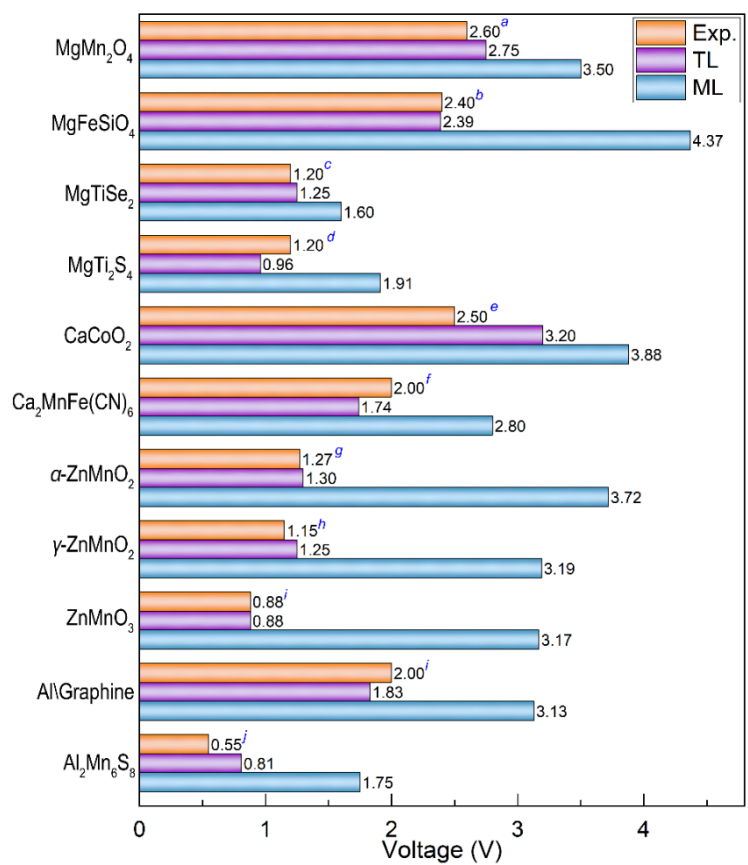

**Fig. 4 | Comparing the predicted voltages of AI models with the available experimental values.**
[a]Ref.[43]. [b]Ref.[44]. [c]Ref.[45]. [d]Ref.[46]. [e]Ref.[47]. [f]Ref.[48]. [g]Ref.[49]. [h]Ref.[50]. [i]Ref.[51]. [j]Ref.[52].

**Online artificial intelligence tool kit for voltage prediction.** A web tool, provided for the voltage prediction for both multivalent and monovalent MIB electrodes, is publicly available online (http://batteries.2dmatpedia.org/). The voltage of a crystal can be obtained within a few seconds, with the crystal structure or the material ID in the MP and type of metal ion as inputs. The voltages for LIBs are predicted by the interpretable DL model and others are predicted by the deep TL model.

**Conclusion**

We develop an interpretable TL model to accurately predict the electrode voltages for MIBs (especially the multivalent batteries with very small data) and explain the underlying physical pictures as the important features for the voltage prediction by visualizing the vectors in the layers in the neural network. The prediction with our model is much accurate in comparison with the ML models and validated by the reported experimental results. The high-performing and interpretable DL models with the booming growth of battery materials data would greatly benefit the battery community that can be used for AI-enabled high-speed materials screening and rational design of electrodes of MIBs, such as substituting high local voltage elements ([Fig. 3c](#)) in a crystal to improve its voltage and energy density.

**Acknowledgments**

The authors thank Dr Zeng Minggang and Dr Xu Lei for their helpful discussion. This work was supported by MOE, Singapore Ministry of Education (No. MOE2019-T2-2-030, No. R-723-000-029-112, and No. R265-000-691-114.), conducted at the National University of Singapore. The authors gratefully acknowledge the Centre of Advanced 2D Materials, National University of Singapore and the National Supercomputing Centre of Singapore for providing computational resources, and the China Scholarship Council for financial support.

**Methods**

**Training Datasets.** The data are extracted from the MP database[32,53] accessed through the application programming interface (API) implemented in pymatgen[54]. The database contains a total number of 4401 intercalation-based battery electrode materials, in which, 2291 entries are electrode materials for LIBs, and 393, 484, 385, 149, 328, and 125 instances for the Mg-, Ca-, Zn-, Al-, Na-, and K-ion battery, respectively (Table S1). Before training, outliers and inconsistent data are removed, and 2190, 387, 471, 378, 149, 287 and 125 instances are kept to train the model for Li-, Mg-, Ca-, Zn-, Al-, Na-, and K-ion battery electrode materials, respectively.

**Machine learning model.** The conventional support vector regression (SVR) and kernel rigid regression (KRR) models are used in this work to predict the electrode voltages. They produce quite similar voltages for both multivalent and monovalent MIBs (Fig. S4). In this work, we refer ML specifically to conventional ML models to distinguish them from the deep learning.

**Deep learning model.** Our DL models are mainly based on the CGCNN.[4] The CGCNN model presents a periodic crystal structure into a multigraph $G$. Each atom in the crystal is represented by a node $i$ in $G$ which is represented by the atom feature vector $v_i$. The vector is then transformed into a 64-dimensional vector in the embedding layer. The nearest 12 neighbors for each atom are also considered in the CGCNN model and the chemical bond between neighbor atoms $i$ and $j$ is expressed as an edge $(i, j)_k$ in $G$ and the bond feature is represented by vector $u_{(i,j)_k}$. The subscript $k$ indicates the $k$-th edge between note $i$ and $j$ because of the periodicity of the crystal[14]. Then, the atom and bond features are input to the convolutional layer using the convolution function designed in the following equation.

$$v_i^{(t)} = v_i^{(t-1)} + \sum_{j,k}[\sigma(z_{(i,j)_k}^{(t-1)}W_f^{(t-1)} + b_f^{(t-1)})\odot g(z_{(i,j)_k}^{(t-1)}W_s^{(t-1)} + b_s^{(t-1)})] \quad (1)$$

where $z_{(i,j)_k}^{(t)} = v_i^{(t)} \oplus v_j^{(t)} \oplus u_{(i,j)_k}$ is a neighbor feature vector, and $\oplus$ denotes the concatenation of the atom and bond feature vectors of the neighboring atoms of the $i$-th atom. $\odot$ denotes element-wise multiplication, $\sigma$ is a sigmoid function, and $g$ is a non-linear activation function. $W$ and $b$ denote weights and biases of the neural networks, respectively. After three convolutional layers, the output vectors are then fitted into a pooling layer to create an overall feature vector. The resulting vector is then fully connected via a hidden layer, followed by a linear transformation to scalar values.

**Interpretability and visualization of deep neural network:** Ante-hoc and post-hoc are the two commonly used interpretation models. Ante-hoc model would open the black box and to gauge how a certain neural network is about to contribute to its predictions. Post-hoc only pries about the model from the outside of the model[55]. In the CGCNN network[4], the output vectors of each layer are available for public, the ante-hoc could be used to probe the whole model. In this work, the element and the local environment representations, who are endured from embedding and the convolutional layers in the model respectively ([Fig. 1](#)), are mapped into two dimensions for visualizing the feature vectors in the layers. The local voltages are also explored to evaluate the contribution of each atom to the crystal voltage.

*Element representations* completely depend on the type of the elements in the crystal. In the CGCNN model, each atom is represented by a node $i$ in the multigraph representation $G$ and the node stores a 92-dimensional vector $v_i$ which include all the elemental information such as the atom covalent radius, the number of valence electrons in each atom orbital. Then, the vector $v_i$ is put into the embedding layer and converted into a 64-dimensional vector $v_i^{(0)}$. In order to visualize the features in the embedding layer, the elemental representation vector $v_i^{(0)}$ are dragged out from the model. But it cannot be directly visualized due to the large dimension of the vector. Thus, we adopt the PCA analysis to reduce the elemental representation into a two-dimensional plane.

*Local environment representations* depend on both the element type and their neighbors in crystal. After the three convolutional layers in the CGCNN model, a 64-dimensional vector $v_i^{(3)}$, who represents the local environments of atom $i$ and contains the information of both the atom and their neighbors, is obtained. In our work, only the vectors with the local oxygen coordination environments, in which the working atom must have at least two oxygen atoms as neighbors, are dragged out for visualization. It is because cooperation among the same environment is more reasonable, and the oxygen coordination environment is the commonest in the electrode materials dataset of MIBs. The local oxygen environment vector $v_i^{(3)}$ is then reduced into two dimensions with the PCA and t-SNE method in order to have a clear visualization about what information has been learned in the convolutional layer.

*Local voltage representation* is derived from the local oxygen coordination environment vectors and represents the contributions of each atom to the voltages of the crystal. A linear transformation is performed to map $v_i^{(3)}$ to the local voltage $V_i$,

$$V_i = v_i^{(3)} W_l + b_l \quad (2)$$

where $W_l$ and $b_l$ denote weights and biases. These two parameter vectors are trained solely during model visualization process. The voltage of the crystal is predicted using the average of the local voltage $V = \frac{1}{n}\sum_i V_i$ where $n$ is the number of atoms in the crystal.

**Transfer learning.** TL is an effective DL model for the prediction with insufficient datasets and is expected to speed up and improve the performance of the convolution neural network. There are two major TL scenarios for loading parameters from previous neural networks. The first method is fine-tuning the convolution network. Herein, the parameters of the target network are initially loaded with a pre-trained network. Thereafter, all the parameters are optimized just as usual. The second method is setting the convolutional network as fixed feature extractor. This method freezes the weights of the earlier layers and only the last few fully connected layers are trained. In this study, the second TL method is used for the voltage predictions of multivalent MIBs and monovalent Na- and K-ion batteries. Only the parameters in the last two fully connected layers in the CNN structure are optimized in the TL model, which are the hidden layer and the output layer (**Fig. 1** orange box).

**Data availability**

All battery data and experimental data are available in Supplemental Materials and GitHub. The data for all figures and extended data figures are available in Source Data.

**Code availability**

Models and results plotting codes are available at

https://github.com/mpeshel/Interpretable_AI_for_battery.git